# Quantifying Effective Noise Sources in Coupled Resonating MEMS sensors

**Vinayak Pachkawade**[*]

[*]**Contact details of the corresponding author**

Vinayak Pachkawade

Email: vinayak.pachkawade@gmail.com

Phone: +32-04-6511 6351

ORCID: vinayak pachkawade

https://orcid.org/0000-0001-7357-3350

## Abstract

This paper presents realistic system-level modelling and simulation of effective noise sources in a coupled resonating MEMS sensors. A governing set of differential equations are used to build a numerical model of a mechanical noise source in a coupled-resonator sensor. An effective thermo-mechanical noise is then quantified through the system-level simulation obtained via Simulink. On a similar note, various noise sources in electronic readout are identified and the contribution of each is quantified to determine an effective noise that stems from the electronic readout. A comparison between an effective mechanical and electronic noise aids in identifying the dominant noise source in a sensor system. A method to optimize the system noise floor for an amplitude-based readout is presented. The proposed models present a variety of operating conditions, such as finite quality factor, varying coupled electric spring strength, and operation with in-phase and out-of-phase mode. The proposed models aim to determine the impact of fundamental noise processes and thus quantify the ultimate detection limit into a coupled resonating system used for various sensing applications.

## 1 Introduction

In a representative MEMS integrated sensor system, electronic circuits such as a current to voltage converter or a charge amplifier is used as a front-end readout [1,2]. These circuits process the electrical signal (for instance, an output motional current from a resonator) provided by the sensing element. A total noise level of the sensor system is therefore due to the combined effect of the mechanical noise of the sensing element, the electrical noise of the (resistive) mechanical sensing element [3] and the input referred noise of the readout circuits [3–6]. In a MEMS sensor system, it is seen that noise from a readout circuit usually dominates the system noise [7]. However, the details of the mechanical noise is also considered to be relevant [8]. Eventually, it is a dominant noise source (mechanical or electronic) that determines the ultimate detection limit of the measurand by the sensing system under consideration.

In this work, we model and quantify an effective noise into the two coupled resonator (CR) system used for sensing applications. Through mathematical analysis and simulations obtained via Simulink model, a source of dominant noise source is identified. Proposed system-level models provides an overview of the efficient method in the sensor design at the system level, taking into account variety of operating conditions. Also, an effective noise floor in the design is quantified. Following this, the finest possible resolution for the sensor is determined that aids in determining the miming detection limit of a quantity to be measured.

## 2 Theory

Consider a schematic representation of a 2-DoF CR sensor system as shown in Figure 1 (a). It consists of effective mass, $M_i$, spring constant, $K_{mi}$ and damping constant, $c_i$, ($i = 1,2$). A coupling spring, $K_c$ and a damping constant, $c_c$ that may exists between the two resonators are also shown. Design conditions are: $M_1 = M_2 = M$, $K_{m1} = K_{m2} = K_m$ and $c_1 = c_2 = c_c = c$, $K_c \ll K_m$. In the design, it is stated that $X_{ji}$ is the amplitude of a $j^{th}$ resonator ($j =1, 2$) at the $i^{th}$ mode of a frequency response ($i =1, 2$) due to the noise induced into the system.

In CR sensor, the change in the sensor output/s due to the change in the input is expressed as follows:

$$\frac{\Delta(\omega_i)_{ji}}{\Delta k} = \frac{1}{2m} \Rightarrow \Delta(\omega_i)_{ji} \approx \left|\frac{\Delta k}{2m}\right| \tag{1}$$

$$\frac{\Delta\left(\frac{x1}{x2}\right)_{ji}}{\Delta k} = \frac{1}{2K_c} \Rightarrow \Delta\left(\frac{x1}{x2}\right)_{ji} \approx \left|\frac{\Delta k}{2K_c}\right| \tag{2}$$

$$\frac{\Delta(x)_{ji}}{\Delta k} = \frac{1}{4K_c} \Rightarrow \Delta(x)_{ji} \approx \left|\frac{\Delta k}{4K_c}\right| \tag{3}$$

Here, $\omega_i$, $(x_1/x_2)_{ji}$ and $x_{ji}$ represent mode-frequency, amplitude ratio and amplitude, respectively. The term $\Delta k$ represents the stiffness perturbations. The normalised output/s for resonant frequency, amplitude ratio (AR), and eigenstate are expressed as

$$\frac{\Delta(\omega_i)_{ji}}{(\omega_i)_{ji}} \approx \left|\frac{\Delta k}{2K_{eff}}\right| \tag{4}$$

$$\frac{\Delta\left(\frac{x1}{x2}\right)_{ji}}{\left(\frac{x1}{x2}\right)_{ji}} \approx \left|\frac{\Delta k}{2K_c}\right| \tag{5}$$

$$\frac{\Delta(x)_{ji}}{(x)_{ji}} \approx \left|\frac{\Delta k}{4K_c}\right| \tag{6}$$

respectively.

In the context of the mass perturbations, $\Delta m$, all forms of the sensor output (normalised) can be derived as follows:

$$\frac{\Delta(\omega_i)_{ji}}{(\omega_i)_{ji}} \approx \left|\frac{\Delta m}{2m_{eff}}\right| \tag{7}$$

$$\frac{\Delta\left(\frac{x1}{x2}\right)_{ji}}{\left(\frac{x1}{x2}\right)_{ji}} \approx \left|\frac{\Delta m}{2K_c}\right| \tag{8}$$

$$\frac{\Delta(x)_{ji}}{(x)_{ji}} \approx \left|\frac{\Delta m}{4K_c}\right| \tag{9}$$

From equations (4) to (9), it is understood that lower value of coupling spring, $K_c$ attain higher relative shifts in the AR and eigenstates (thus sensitivity) for a given perturbation, $\Delta k$ ($\Delta m$). Equations (1) to (3) represents a minimum resolvable shifts (due to $\Delta k$) into the frequency, amplitude ratio (AR), and eigenstate, respectively. In this work, we focus on the amplitude based output and amplitude noise.

## 3   Thermo-mechanical noise in CR sensor

In order to quantify the effective system noise in a coupled resonator (CR) sensor, a noise source was added into the model. An output response was analysed to capture the sensor output resolution. This procedure aided to determine the ultimate detection limit of a physical measurand. In this section, we model and quantify a magnitude of a thermal-mechanical noise in the design. Micro-nano size moving parts in MEMS are especially susceptible to the mechanical noise resulting from molecular agitation [3]. In ultra-low level signal detection, mechanical-thermal noise plays an important role in setting up the effective noise floor of a sensor system, and, thus a minimum detection limit. In the context of a CR sensor, it has been postulated that it is the thermal-mechanical noise that governs the ultimate detection limit of a CR sensor [4]. In this paper, a mechanical-thermal noise source in 2 DoF CR sensor is modelled. Its impact on the output signal resolution (i.e. lowest possible detectable physical quantity) is quantified for the amplitude based output of the design.

To validate the theoretical model as derived above, a power spectral density (PSD) of an amplitude displacement noise of $j^{th}$ resonator ($j = 1, 2$) at the $i^{th}$ mode of the frequency response ($i = 1, 2$) was obtained through the Simulink model. It is shown in Figure 1(b). A mechanical noise source was modelled by adding a force term into the governing set of equations of motion (for 2 DoF) as follows

$$M\ddot{X}_1 + (c+c_c)\dot{X}_1 + (K_m + K_c)X_1 - c\dot{X}_2 - K_c X_2 = F_{noise\_rms} \tag{10}$$

$$M\ddot{X}_1 + (c+c_c)\dot{X}_1 + (K_m + K_c)X_1 - c\dot{X}_2 - K_c X_2 = F_{noise\_rms} \tag{11}$$

A spectral density of a noise forcing term is given as $F_{noise\_density} = \sqrt{4\,k_B T\,c}$ $N/Hz^{0.5}$ [3,8]. Here, $k_B$ is the Boltzmann constant ($\approx 1.380\times10^{-23}$ *Joule/Kelvin*), $T$ is the temperature (300 *Kelvin*) and $c$ is the damping coefficient ($c = 0.0031$ *Ns/m* in our design). As observed, the spectral density of a mechanical noise force depends on temperature and the magnitude of mechanical damping. A noise forcing term with the calculated average value, $5.136 \times 10^{-22}$ $N^2$ was added as an excitation force in to the Simulink model.

A noise PSD (in *dB/Hz*) for displacements $X_1$ and $X_2$ was plotted as shown in Figure 2. Simulations were run for the varying strength of a coupling spring, $K_c$ between the two resonators. Condition for the simulation are as follows: $c$ = 0.0031 *Ns/m*, ($Q \approx$ 2547), $\Delta k$ = 0, $\kappa$ = -0.0032, excitation force was applied to mass $M_1$. As seen from refer Figure 2 (a), a resultant displacement noise PSD (due to noise forcing term) is -582.2 *dB/Hz* (at mode 1) and -572.2 *dB/Hz* (at mode 2). These PSD values correspond to an equivalent magnitude $|X_j|$ of 7.76×10$^{-30}$ *m²/Hz* (at mode 1) and 2.45×10$^{-29}$ *m²/Hz* (at mode 2) respectively. Assuming a measurement bandwidth of 10 *Hz* around the resonant mode frequencies, $\omega_i$ ($i$ = 1, 2), an average (mean square) value of a mechanical thermal noise of $j^{th}$ resonator at the $i^{th}$ mode ($j, i$ = 1, 2) is determined as $X_{ji\_avg} = |X_{ji}| \times df$. As observed from Figure 2, noise magnitude is lower for mode 1. Subsequently, a theoretical estimation of effective mechanical-thermal noise current for our design is $i_{motX_{ji}} = \eta \omega_i X_{ji}$, where, $\eta$, $\omega_i$, and $X_{ji}$ are transduction factor, angular frequency and the maximum displacement amplitude of the $j^{th}$ resonator ($j$=1,2) at the $i^{th}$ mode of the frequency response ($i$=1,2), respectively [7]. By determining the $X_{ji}$, the corresponding motional currents, $i_{motX_{ji}}$ was quantified. Since, a thermal noise amplitude of the mode 1 is relatively lower (referring Figure 2), the computation for the best case is given in Table 1. Therefore, effective mechanical-thermal noise floor is $\approx 4.9 \times 10^{-15} A_{rms}$.

A value of $i_{motX_{j1}}$ signifies the lower limit of detecting the change in the CR sensor output motion current due to the change in the input perturbation to be measured ($\Delta k$ or $\Delta m$). Moreover, a higher quality factor, Q can mitigate the effect of mechanical-thermal noise and thereby further improve this detection limit. In our design, it is determined that any variation in the output modal amplitudes of the $j^{th}$ resonator ($j$ = 1, 2) is resolvable as long as the simulated/measured individual *rms* amplitude shifts, $x_{ji}(i_{mot_{ji}})$ is greater than the *rms* amplitude fluctuations, $X_{ji}(i_{motX_{ji}})$ induced by the noise in the system, *i.e. SNR* $\geq$ 1 (referring to the value of $i_{motX_{j1}}$).

## 4 Electronic readout noise

Figure 3 (a) represents the schematic representation of current-to-voltage converter that may be implemented using OPA 381 [9] integrated circuit (IC). This configuration acts as a readout circuit to the CR sensor. Two such ICs may be used into each output channel of the $j^{th}$ resonator ($j$ = 1, 2) of a CR sensor for motion current pick-up. Figure 3 (b) is the derived equivalent noise model for the readout circuit. By superposition and nodal analysis, we estimated noise contributions from each of these sources as follows.

An amplifier (OPA 381) [9] used in the analysis has an input current noise density, $i_n \approx$ 20 *fA/Hz*$^{0.5}$, and an input voltage noise density, $v_n \approx$ 70 *nV/Hz*$^{0.5}$. To determine the lowest possible resolvable shifts into the amplitude based output/s (and thus quantify input perturbations, $\Delta k$ or $\Delta m$), all the noise components of the readout circuit are re-expressed as an equivalent currents as given in Table 2. In Table 2, $i_j^{noise}$ (*rms*) is the effective noise current from the $j^{th}$ resonator at the $i^{th}$ mode of response. $R_x$ is the motional resistance, $R_x = \dfrac{d^4 \sqrt{k_{eff} M}}{V_{dc}^2 \varepsilon^2 A^2 Q}$ [10] ($R_x$ = 4 *MΩ* for a $Q \approx$ 2547 with other parameters being constant, in our case), $R_f$ is the effective value of a feedback resistor used in the circuit (1 *MΩ*) (refer Figure 3). The term $k_B$ is the Boltzmann constant ($\approx$ 1.380×10$^{-23}$ *Joule/Kelvin*) and *T* is the temperature (300 *Kelvin*). Term *B* is the integration bandwidth of 10 *Hz* around the resonator's mode frequencies, $f\_{ip}$ and $f\_{op}$ used in the computations for optimum noise estimation. From Table 2, an effective theoretical noise floor (due to electronic readout) of our design is $\approx 1.56 \times 10^{-13}$ $A_{rms}$. By comparision, we can see that it is the mechanical-thermal noise current that sets the ultimate noise floor. However, it is the input-reffered noise current from the electronic readout that dominates the noise performance of the integrated sensor system. Consequently, the total system noise (mechanical sensor + electronic readout) in our design is then estimated by vector sum of the uncorrelated noise sources i.e. $I_t^2 = I_1^2 + I_2^2$, where, $I_1$ is effective mechanical noise current and $I_2$ is effective electronic readout noise current. $I_t$ is calculated to be $\approx 1.56 \times 10^{-13}$ $A_{rms}$.

Figure 4 shows a system-level Simulink model for the complete CR sensor system. In this model, the electrostatic actuation force is applied to mass 1. Here, other operating conditions that were used to obtain results in Figure 2 are same. Output motion current of the $j^{th}$ resonator undergoes a voltage amplification by the factor 1MΩ. The total noise current, $I_t$ as estimated above also undergoes the same amplification factor to yield output referred noise voltage $V_{ji}^{noise}$. As shown in Figure 4, corresponding output voltages, $v_{out_{ji}}$ of the sensor system can be recorded. Therefore, in the presence of an applied electrostatic force at the $i^{th}$ mode frequency, impact of system noise on the motional current output, $i_{mot_{ji}}$ or voltage output, $v_{out_{ji}}$ of $j^{th}$ resonator ($j$=1, 2) at the $i^{th}$ mode of a response ($i$=1, 2) can be determined.

## 5 Sensor output resolution

Table 3 provides the corresponding computations and simulated results. The calculated values for motional current amplitudes are in agreement with the simulated values. The minimum resolvable shift in the voltage amplitudes of our 2-DoF CR sensor was derived as $\approx \dfrac{V_{ji}^{noise}}{V_{out_{ji}}}$ [7]. Here, $V_{ji}^{noise}$ is the output refereed noise voltage of the $j^{th}$ coupled resonator at the $i^{th}$ mode. The term $V_{out_{ji}}$ is the noiseless deterministic output voltages (determined from the theoretical expression) given as $V_{out_{ji}} = i_{mot_{ji}} \times R_f$, where $i_{mot_{ji}} = \eta \omega_i x_{ji}$. Note that $V_{out_{ji}}$ is also available from the Simulink model and, theory and simulations results match. Moreover, normalized values of the sensor outputs can be expressed by the expression $\dfrac{\Delta V_{out_{ji}}}{V_{out_{ji}}}$, where, $\Delta V_{out_{ji}}$ is the lowest possible resolvable mode amplitude shift of $j^{th}$ resonator ($j$=1, 2) at the $i^{th}$ mode of a response ($i$=1, 2) due to the applied perturbation. Theoretically, for the $j^{th}$ resonator, such amplitude shift for mode 1, can be obtained as $\dfrac{V_{j1}^{noise}}{V_{out_{j1}}} \approx \dfrac{156 \times 10^{-9}}{0.135} = 1.15 \times 10^{-6}$ (dimensionless). Also, for mode 2, lowest possible resolvable amplitude shift is $\dfrac{V_{j2}^{noise}}{V_{out_{j2}}} \approx \dfrac{156 \times 10^{-9}}{0.271} = 5.756 \times 10^{-7}$ (dimensionless). Similarly, minimum resolvable shift in the amplitude ratio readout, $AR_i$ at the $i^{th}$ mode ($i$=1, 2) $\sqrt{\left(\dfrac{V_{11}^{noise}}{V_{out_{11}}}\right)^2 + \left(\dfrac{V_{21}^{noise}}{V_{out_{21}}}\right)^2} \approx 1.481 \times 10^{-6}$ (dimensionless) for mode 1 and $\sqrt{\left(\dfrac{V_{12}^{noise}}{V_{out_{12}}}\right)^2 + \left(\dfrac{V_{22}^{noise}}{V_{out_{22}}}\right)^2} \approx 3.89 \times 10^{-7}$ (dimensionless) for mode 2. Here, $V_{ji}^{noise}$ and $V_{out_{ji}}$ are the corresponding output referred noise and output voltages, respectively of the $j^{th}$ CR at their associated $i^{th}$ mode of the frequency response. Since rms amplitudes of output voltage (essentially a motional current) at the mode 1 for $j^{th}$ resonator is relatively closer (worst-case signal-to-noise (S/N) ratio) to the rms noise amplitude, it gives the possibility to determine the worst-case lowest possible shift (i.e. sensor output resolution). The effective resolution (theoretical) of our design for AR based readout is $\approx 3.89 \times 10^{-7}$. This value may be compared to the theoretically calculated resolution ($\approx 6 \times 10^{-3}$) in MEMS coupled resonator design in [7].

## 5.1 Sensing system and noise

Figure 5 illustrates a power spectrum of our design for the motion current outputs. Operating condition are $c = 0.0031$ $Ns/m$, ($Q \approx 2547$), $\Delta k = 0$, $\kappa = -0.0032$, excitation force F $\approx 149$ $\mu N$ (peak-to-peak) applied to mass $M_1$. Sensor output signal power levels at the resonant frequencies are determined with and without the effective noise added to the system. A value of an effective noise power used in the Simulink block is $\approx 1 \times 10^{-26}$ W (Figure 5 (a)-(b)) and $\approx 1 \times 10^{-24}$ W (Figure 5 (c)-(d)). Figure 6 presents power spectral density (PSD) curves (left Y-axis) of an effective output referred noise power and sensor output signal power for $j^{th}$ resonator at the mode 1 (Figure 6 (a)) and mode 2 (Figure 6(b)) of the frequency response, respectively. MATLAB was used to estimate the signal's average power by "integrating" under its PSD curve. A calculated average noise power is $\approx 1 \times 10^{-7}$ $W$ (equivalent to effective output referred noise voltage of $3.16 \times 10^{-4}$ Volts). Note that, actual estimated noise floor (both in theory and simulation) is $\approx 1 \times 10^{-13}$ Amp ($1 \times 10^{-26}$ W), which, post amplification (by factor 1 $M\Omega$), provides an output referred noise voltage $\approx 1 \times 10^{-7}$ Volts ($1 \times 10^{-14}$ W). In this case, a calculated average signal power is $\approx 0.2$ Volts (0.07 W) for resonator 1 and 0.254 Volts (0.064 W) for resonator 2. All our calculations are for an assumed reference load of 1$\Omega$.

## 5.2 Quantifying minimum detectable shift in the measurand

For a proposed design, AR based sensitivity to the applied stiffness perturbation is determined to be 180 V/V/$\delta_k$. Here, $\delta_k$ is the normalized stiffness perturbation, expressed as $\Delta k/K_{eff}$. Also, as given in section 5 above, sensor's AR based output resolution is determined to be $3.89 \times 10^{-7}$ (dimensionless). Therefore, a minimum detectable stiffness is obtained as follows:

$$\text{Min. detectable stiffness } (N/m) = \frac{AR \text{ output resolution } (dimensionless)}{Sensitivity \ (V/V/N/m)}$$

$\text{Min. detectable stiffness } (N/m) = \frac{3.89 \times 10^{-7}}{180/\delta_k} = 2.161 nN/m$. Assuming an integration bandwidth of 10 Hz, around the resonant mode-frequency of interest in the 2-DoF CR sensor, spectral density of stiffness perturbation is obtained as

$$\text{Min. detectable stiffness } (N/Hz^{0.5}) = \frac{AR \text{ output resolution } (V/V/Hz^{0.5})}{Sensitivity \ (V/V/\delta_k)}$$

$\text{Min. detectable stiffness } = \frac{1.23 \times 10^{-7} / Hz^{0.5}}{3.615 / N/m} = 0.68 nN/m/Hz^{0.5}$

## 6 Influence of coupling strength on the effective noise floor

A strength of electrostatic spring, coupling the two resonators influence the effective noise floor in the sensing system. For the analysis, we considered the coupling factor, $\kappa$ that separates two modes by about 7 Hz. A quality factor, $Q$ of about 2547 was used in this design. Figure 7 shows a simulated spectrum of both the resonators for applied electrostatic forcing term. As observed, with the given set of operating conditions, motional current output signal from resonator 2 offers relatively lower noise floor as compared to resonator 1 and is independent of the coupling strength. In addition, it is beneficial to utilize output of $j^{th}$ resonator at mode 2 as it offers a higher amplitude and thus maximum shifts (sensitivity). From Figure 8, we can infer by comparison that smaller coupling strength in the design leads to reduced noise floor (about -12 *dB* gain in noise floor) for the resonating output of the $j^{th}$ resonator ($j = 1, 2$).

## 7 Conclusion

This paper quantitatively and qualitatively evaluated the impact of two main noise processes in the emerging class of a CR sensors. For the first-time, realisitic and detailed-oriented system-level behavioral models are built and presented to monitor the output response of CR sensors. Such models are currently not available in the existing literature. The proposed models quantify effective mechanical-thermal and electronic noise and aids in determine the dominant noise source in the sensor system. Furthermore, by determining the sensor output resolution (essentially effective noise floor) and the parametric sensitivity, we are able to estimate the lowest possible detectable quantity under measurement.

**Funding**

This research did not receive any specific grant from funding agencies in the public, commercial, or not-for-profit sectors.

**Figure captions**

**Figure 1.** A schematic representation of a 2-DoF coupled MEMS resonator used as a sensor (a) and its equivalent system-level behavioral model representation in Simulink (b)

**Figure 2.** Simulated power spectrum density (PSD) of a displacement noise of resonator 1 and 2 subject to mechanical-thermal noise forcing term applied on resonator 1. (a) $K_c$ = -393.5 N/m and (b) $K_c$ = -1000 N/m. Other simulation conditions are as follows: $c$ = 0.0031 Ns/m, ($Q \approx 2547$), $\Delta k$ = 0, $\kappa$ = -0.0032, excitation force applied to mass $M_1$. A normalized coupling factor, $\kappa$ is the given by $K_c/K_{eff}$. $\Delta k$ is the induced perturbation into the stiffness of one of the resonator in 2-DoF CR sensor system.

**Figure 3.** (a) OPA 381 interface for a 2-DoF CR sensor design (b) equivalent noise circuit to evaluate circuit dominant noise.

**Figure 4.** A system-level model to evaluate the performance of a CR sensor system

**Figure 5.** Simulated power spectrum of our design across the frequency range of interest, (a)-(b) with less noise current, (c)-(d) with increased noise current

**Figure 6.** Simulated PSD and power spectrum plot: For an effective output referred noise (left y-axis) and signal power output (right y-axis) for (a) mode 1 and (b) mode 2 in our two scaled-up weakly coupled 2 DoF architecture

**Figure 7.** Simulated power spectrum density (PSD) of the motion current output signal in a two weakly coupled resonator for the following operating condition, $Q$ = 2547, $\kappa$ = -0032, $\delta_k$ = 0, $F$ =1 N

**Figure 8.** Simulated power spectrum density (PSD) of the motional current output signal for varying coupling strength, $K_c$. electrostatic drive forcing term set to unity. For smaller coupling, Kc = -393.5 N/m, output from resonator 2 can be utilized as it provides the lower effective noise floor.

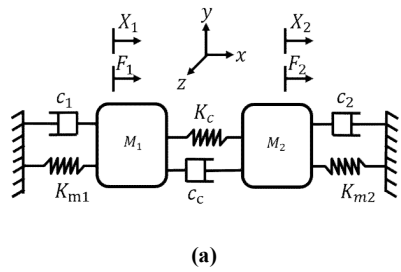
(a)

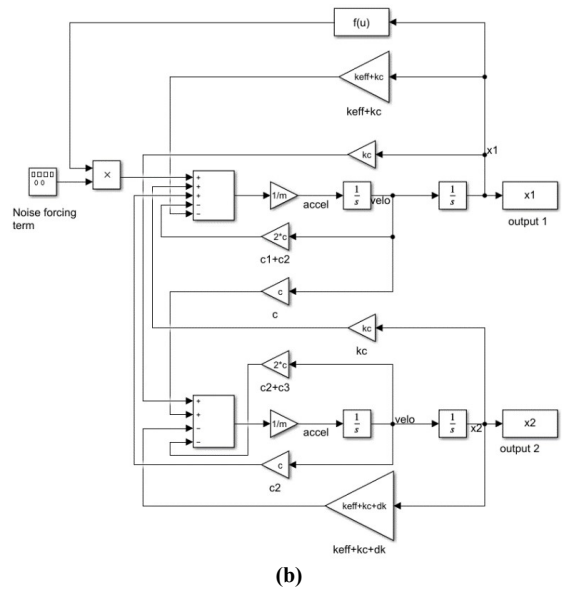
(b)

Fig. 1

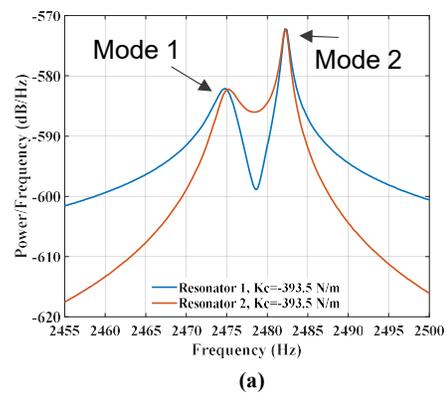 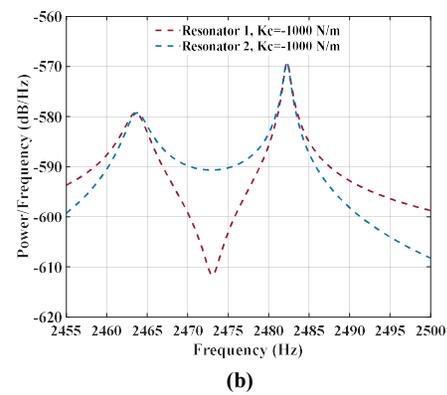

Fig. 2

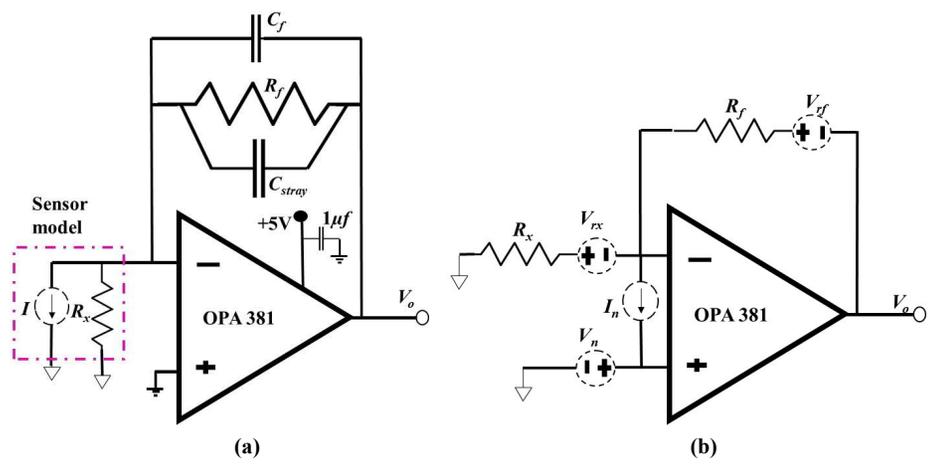

Fig. 3

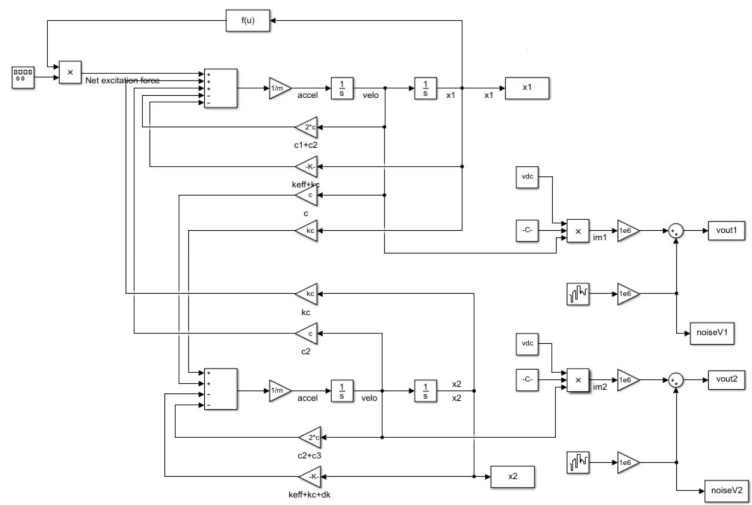

Fig. 4

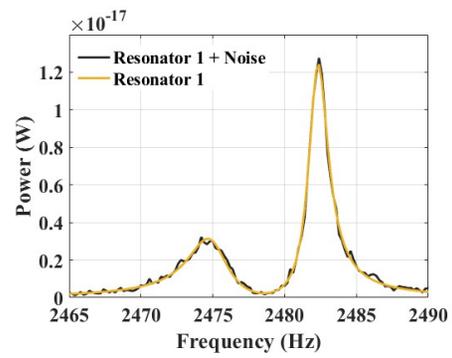 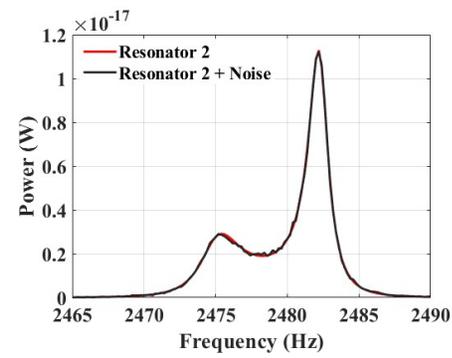
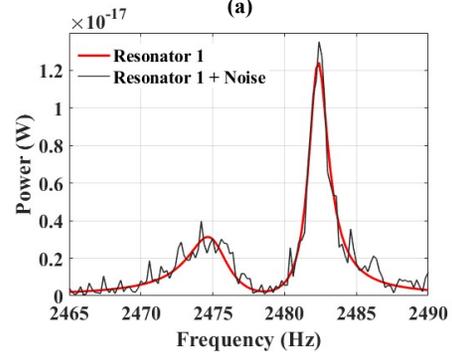 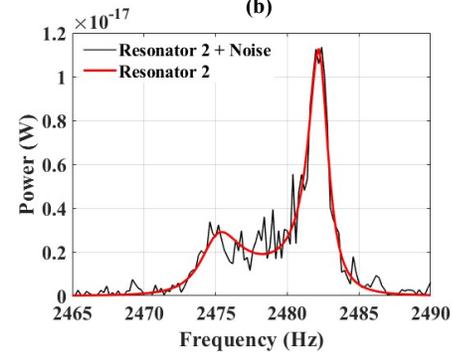

Fig. 5

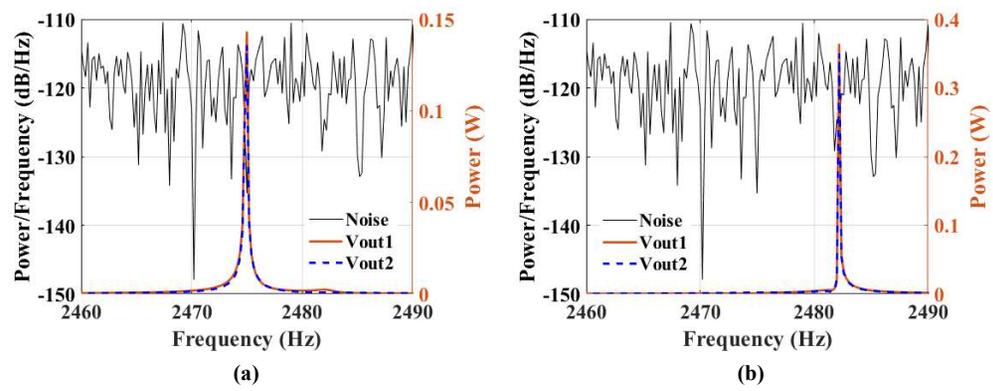

Fig. 6

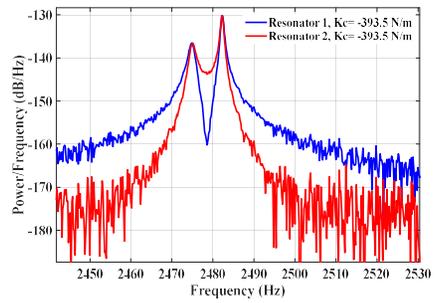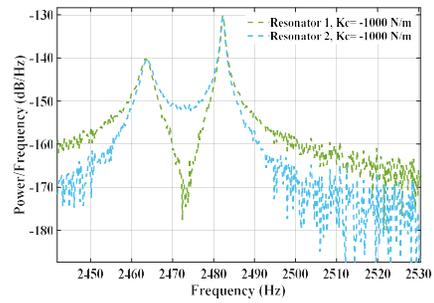

Fig. 7

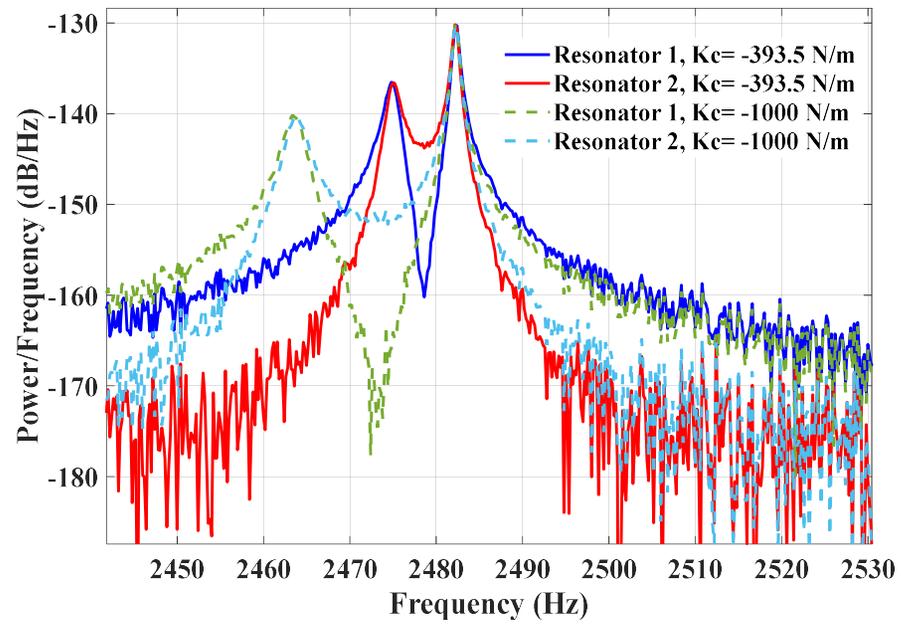

Fig. 8

**Tables:**

**Table 1.** Table showing computation of the effective noise forcing term, output displacement noise amplitudes and the resultant motional noise current.

| Term | Value | Expression |
|---|---|---|
| $X_{j1\_avg}$ | $\approx 7.762 \times 10^{-29} \, m^2$ | $X_{ji\_avg} = |X_{ji}| \times df$ [1] |
| $X_{ji\_rms}$ | $\approx 8.81 \times 10^{-15} \, m$ | $X_{j1\_rms} = \sqrt{X_{j1\_avg}}$ [2] |
| $i_{motX_{j1}}$ | $\approx 4.9 \times 10^{-15} \, A_{rms}$ | $i_{motX_{ji}} = \eta \omega_i X_{ji}$ [3] |
| $F_{noise\_PSD}$ | $5.136 \times 10^{-23} \, N^2/Hz$ | $\left(F_{noise\_density}\right)^2 = 4 k_B T \, c$ [*] |
| $F_{noise\_avg}$ | $5.136 \times 10^{-22} \, N^2$ | $F_{noise\_avg} = \int_0^{10} \left(\sqrt{4 k_B T \, c}\right)^2 df$ [**] |
| $F_{noise\_rms}$ | $2.26 \times 10^{-11} \, N$ | $F_{noise\_rms} = \sqrt{F_{noise\_avg}}$ [***] |

[1] A best-case estimate for average noise power for mode 1. [2] An effective *rms* value of a mechanical-thermal noise for mode 1. [3] An effective noise current due to the mechanical noise forcing term. [*] Power spectral density (PSD) of mechanical noise force generator. [**] Average (mean square) value of a mechanical thermal noise force generator, assumed measurement bandwidth is 10 *Hz*. [***] A effective *rms* amplitude of mechanical noise forcing term

**Table 2.** Table showing computation of the effective noise current due to several noise terms present in the electronic readout used for the mechanical CR sensor.

| Noise sources | Effective noise current ($A_{rms}$) | Expression |
|---|---|---|
| Mechanical thermal-noise [1] | $4.9 \times 10^{-15}$ | $i_{X_{ji}} = \eta \omega_i X_{ji}$ |
| Feedback resistance | $4.06 \times 10^{-13}$ | $\sqrt{\dfrac{4 k_B T \times B}{R_f}}$ |
| Input voltage noise of a pre-amplifier | $4.34 \times 10^{-13}$ | $\sqrt{v_n^2 \left(1+\dfrac{R_x}{R_f}\right)^2 / R_x^2} \times \sqrt{B} \times 1.57$ |
| Input current noise of a pre-amplifier [2] | $9.92 \times 10^{-14}$ | $\sqrt{i_n^2} \times \sqrt{B} \times 1.57$ |
| input-referred electronic noise current, $i_j^{noise}$ | $1.56 \times 10^{-13} \, A_{rms}$ | $i_j^{noise} = \sqrt{\left(i_j^{noise}\right)^2} = \sqrt{\left[i_n^2 + v_n^2\left(1+\dfrac{R_x}{R_f}\right)^2 / R_x^2 + \dfrac{4 k_B T}{R_f}\right]}$ |
| Total system noise (mechanical + electronic) | $\approx 1.56 \times 10^{-13} \, A_{rms}$ | $I_t^2 = I_1^2 + I_2^2$ |

[1] best-case calculation for the out-of-phase mode as it shows the lower noise amplitude. [2] factor 1.57 is the roll-off rate of a filter (1-pole) [9].

**Table 3.** Resultant computations and simulated results in response to the electrostatic forcing term used in the Simulink model. A noise estimation due to mechanical and electronic terms were taken into account to simulate for the output response.

| Term | Value (mode 1) | Value (mode 2) | Expression |
|---|---|---|---|
| maximum displacement [1], $\lvert x_{ji} \rvert$ | $x_{j1} \approx 0.419\ \mu m$ | $x_{j2} \approx 0.836\ \mu m$ | NA |
| motional current amplitudes [2], $i_{mot_{ji}}$ | $\approx 192\,nA$ | $\approx 384\,nA$ | $i_{mot_{j1}} \approx \eta \times 2\pi f_{\_op} \times x_{j1}$, $i_{mot_{j2}} \approx \eta \times 2\pi f_{\_ip} \times x_{j2}$ |
| Effective output voltage [3], $V_{out_{ji}}$ | $= 0.135 V_{rms}$ | $= 0.271 V_{rms}$ | $V_{out_{j1}} = i_{mot_{j1}} \times R_f \approx 0.192 V_{max}$, $V_{out_{j2}} = i_{mot_{j2}} \times R_f \approx 0.384 V_{max}$ |
| output referred noise voltage [2], $V_{ji}^{noise}$ | $\approx 156 \times 10^{-9}\ V_{rms}$ | NA | $i_{j1}^{noise} \times R_f$ |
| Sensor output resolution for amplitude readout, $\approx \dfrac{V_{ji}^{noise}}{V_{out_{ji}}}$ | $\approx \dfrac{156 \times 10^{-9}}{0.135} \approx 1.15 \times 10^{-6}$ (dimensionless) | $\approx \dfrac{156 \times 10^{-9}}{0.271} \approx 5.756 \times 10^{-7}$ (dimensionless) | $\dfrac{V_{j1}^{noise}}{V_{out_{j1}}}, \dfrac{V_{j2}^{noise}}{V_{out_{j2}}}$ |
| Sensor output resolution for amplitude ratio $AR_i$ readout, | $\approx 1.481 \times 10^{-6}$ (dimensionless) | $\approx 3.89 \times 10^{-7}$ (dimensionless) | $\sqrt{\left(\dfrac{V_{11}^{noise}}{V_{out_{11}}}\right)^2 + \left(\dfrac{V_{21}^{noise}}{V_{out_{21}}}\right)^2}$, $\sqrt{\left(\dfrac{V_{12}^{noise}}{V_{out_{12}}}\right)^2 + \left(\dfrac{V_{22}^{noise}}{V_{out_{22}}}\right)^2}$ |

[1] Derived from the Simulink. [2] Theory. [3] amplification factor, $R_f = 1\ M\Omega$, as a feedback resistor in preamplifier OPA 381.